\documentstyle[12pt,fullpage]{article}
\def\be{\begin{equation}}
\def\ee{\end{equation}}
\def\beq{\begin{eqnarray}}
\def\eeq{\end{eqnarray}}
\def\lsim{\:\raisebox{-0.5ex}{$\stackrel{\textstyle<}{\sim}$}\:}
\def\gsim{\:\raisebox{-0.5ex}{$\stackrel{\textstyle>}{\sim}$}\:} 
\begin{document}
\begin{flushright}
TIFR/TH/99-59 \\
\end{flushright}
\bigskip
\begin{center}
{\bf Basic Constituents of Matter and their Interactions -- A Progress
Report} \\[2cm]
D.P. Roy \\[1cm]
Department of Theoretical Physic \\
Tata Institute of Fundamental Research \\
Homi Bhabha Road, Mumbai - 400 005, India 
\end{center}
\bigskip\bigskip

Our concept of the basic constituents of matter has undergone two
revolutionary changes -- from atoms to proton \& neutron and then onto
quarks \& leptons.  Indeed all these quarks and leptons have been
seen by now along with the carriers of their interactions, the gauge
bosons.  But the story is not complete yet.  A consistent theory of
mass requires the presence of Higgs bosons along with SUSY particles,
which are yet to be seen.  This is a turn of the century account of
what has been achieved so far and what lies ahead.

\newpage

\noindent \underbar{\bf Introduction}
\medskip

Our concept of the basic constituents of matter has undergone two
revolutionary changes during this century.  The first was the
Rutherford scattering experiment of 1911, bombarding $\alpha$
particles on the Gold atom.  While most of them passed through
straight, occasionally a few were deflected at very large angles.
This was like shooting bullets at a hay stack and finding that
occasionally one would be deflected at a large angle and hit a
bystander or in Rutherford's own words ``deflected back and hit you on
the head''!  This would mean that there is a hard compact object
inside the hay stack.  Likewise the Rutherford scattering experiment
showed the atom to consist of a hard compact nucleus, serrounded by a
cloud of electrons.  The nucleus was found later to be made up of
protons and neutrons.

The second was the electron-proton scattering experiment of 1968 at
the Stanford Linear Accelerator Centre, which was awarded the Nobel
Prize in 1990.  This was essentially a repeat of the Rutherford
scattering type experiment, but at a much higher energy.  The result
was also similar as illustrated below.  It was again clear from the
pattern of large angle scattering that the proton is itself made up of
three compact objects called quarks.

\vspace{6cm}
\includegraphics{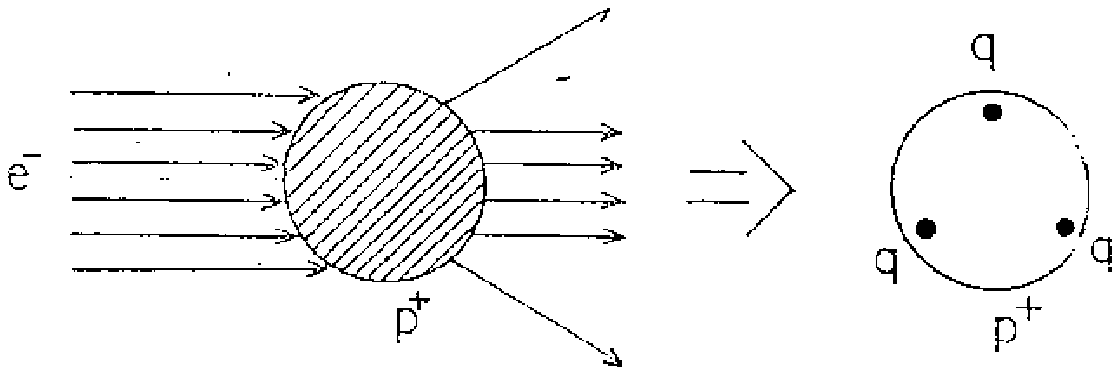} 
\label{fig:9959fig1}
\begin{center}
Fig. 1.  The SLAC electron-proton scattering experiment, revealing the
quark structure of the proton.
\end{center}

\noindent We now know from many such experiments that the nuclear
particles (proton, neutron and mesons, which are collectively called
hadrons) are all made up of quarks -- i.e. they are all quark atoms. 

The main difference between the two experiments comes from the fact
that, while the dimension of the atom is typically $1 A^\circ =
10^{-10} m$ that of the proton is about $1 fm$ (fermi or femtometer)
$= 10^{-15} ~{\rm m}$.  It follows from the Uncertainty Principle,
\be
\Delta E \cdot \Delta x > \hbar c \sim 0.2 ~{\rm GeV.fm},
\label{one}
\ee 
that the smaller the distance you want to probe the higher must be
the beam energy.  Thus probing inside the proton $(x \ll 1 fm)$
requires a beam energy $E \gg 1 ~{\rm GeV} (10^9 ~{\rm eV})$, which is
the energy acquired by the electron on passing through a billion
(Gega) volts.  It is this multi-GeV acceleration technique that
accounts for the half a century gap between the two experiments.

Following the standard practice in this field we shall be using the
so-called natural units,
\be
\hbar = c = 1,
\label{two}
\ee
so that the mass of particle is same as its rest mass energy
$(mc^2)$.  The proton and the electron masses are
\be
m_p \sim 1 ~{\rm GeV}, ~m_e \sim 1/2 ~{\rm MeV}.
\label{three}
\ee
The GeV is commonly used as the basic unit of mass, energy and
momentum. 
\bigskip

\noindent \underbar{\bf The Standard Model}:
\medskip

As per our present understanding the basic constituents of matter are
a dozen of spin-1/2 particles (fermions) along with their
antiparticles.  These are the three pairs of leptons (electron, muon,
tau and their associated neutrinos) and three pairs of quarks (up,
down, strange, charm, bottom and top) as shown below.  The masses of
the heaviest members are shown paranthetically in GeV units.

\begin{center}
Basic Constituents of Matter
\end{center}
\[
\begin{tabular}{|lcccc|}
\hline
&&&& \\
& $\nu_e$ & $\nu_\mu$ & $\nu_\tau$ & 0 \\
leptons &&&& \\
& $e$ & $\mu$ & $\tau(2)$ & -1 \\
&&&& \\
\hline
&&&& \\
& $u$ & $c$ & $t(175)$ & $2/3$ \\
quark &&&& \\
& $d$ & $s$ & $b(5)$ & $-1/3$ \\
&&&& \\
\hline
\end{tabular}
\]

\noindent The members of each pair differ by 1 unit of electric charge
as shown in the last column -- i.e. charge 0 and -1 for the neutrinos
and charged leptons and 2/3 and -1/3 for the upper and lower quarks.
This is relevant for their weak interaction.  Apart from this electric
charge the quarks also possess a new kind of charge called colour
charge.  This is relevant for their strong interaction, which binds
them together inside the nuclear particles (hadrons).

There are four basic interactions among these particles -- strong,
electromagnetic, weak and gravitational.  Apart from gravitation,
which is too weak to have any practical effect on their interaction,
the other three are all gauge interactions.  They are all mediated by
spin 1 (vector) particles called gauge bosons, whose interactions are
completely specified by the corresponding gauge groups. 
\newpage
\begin{center}
Basic Interactions
\end{center}
\[
\begin{tabular}{|l|lll|}
\hline
&&& \\
Interaction & Strong & EM & Weak \\
&&& \\
Carrier & $g$ & $\gamma$ & $W^\pm \& Z^0$ \\
&&& \\
Gauge Group & $SU(3)$ & \multicolumn{2}{c|}{$\underbrace{U(1)~~ SU(2)}$} \\
&&& \\
\hline
\end{tabular}
\]

\noindent Each of the three interactions is represented below by the
corresponding Feynman diagram, which is simply a space-time picture of
scattering with time running vertically upwards.  The 4-momentum
squared transfered between the particles is denoted by $Q^2$, which is
a Lorentz invariant quantity.  The corresponding scattering
amplitudes, representing the square-roots of the scattering
cross-sections (rates), are denoted below each diagram.

\vspace{7cm}
\includegraphics{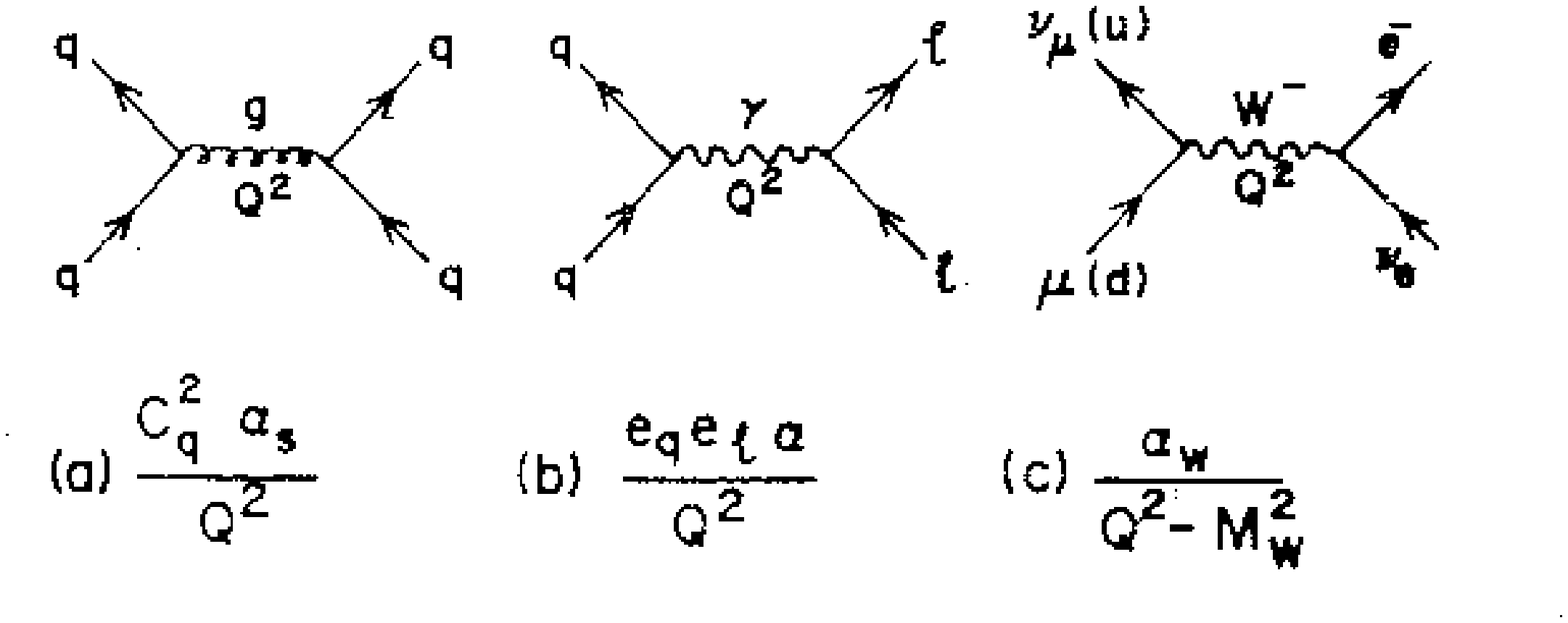} 
\label{fig:9959fig2}
\begin{center}
Fig. 2. The scattering diagrams and amplitudes for (a) strong, (b)
electromagnetic and (c) weak interactions.
\end{center}

\noindent The strong interaction between quarks is mediated by the
exchange of a massless vector boson called gluon.  This is analogous
to the photon, which mediates the electromagnetic interaction between
charged particles (quarks or charged leptons).  The gluon coupling is
proportional to the colour charge just like the photon coupling is
proportional to the electric charge.  The constant of proportionality
for the strong interaction is denoted by $\alpha_s$ in analogy with
the fine structure constant $\alpha$ in the EM case.  And the theory
of strong interaction is called quantum chromodynamics (QCD) in
analogy with the quantum electrodynamics (QED).  The major difference
of QCD with respect to the QED arises from the nonabelian nature of
its gauge group, $SU(3)$.  This essentially means that unlike the
electric charge the colour charge can take three possible directions
in an abstract space.  These are rather whimsically labelled red, blue
and yelow as illustrated below.  Of course the cancellation of the
colour charges of quarks ensure that the nuclear particles (hadrons)
composed of them are colour neutral just like the atoms are
electrically neutral.

\vspace{4cm}
\includegraphics{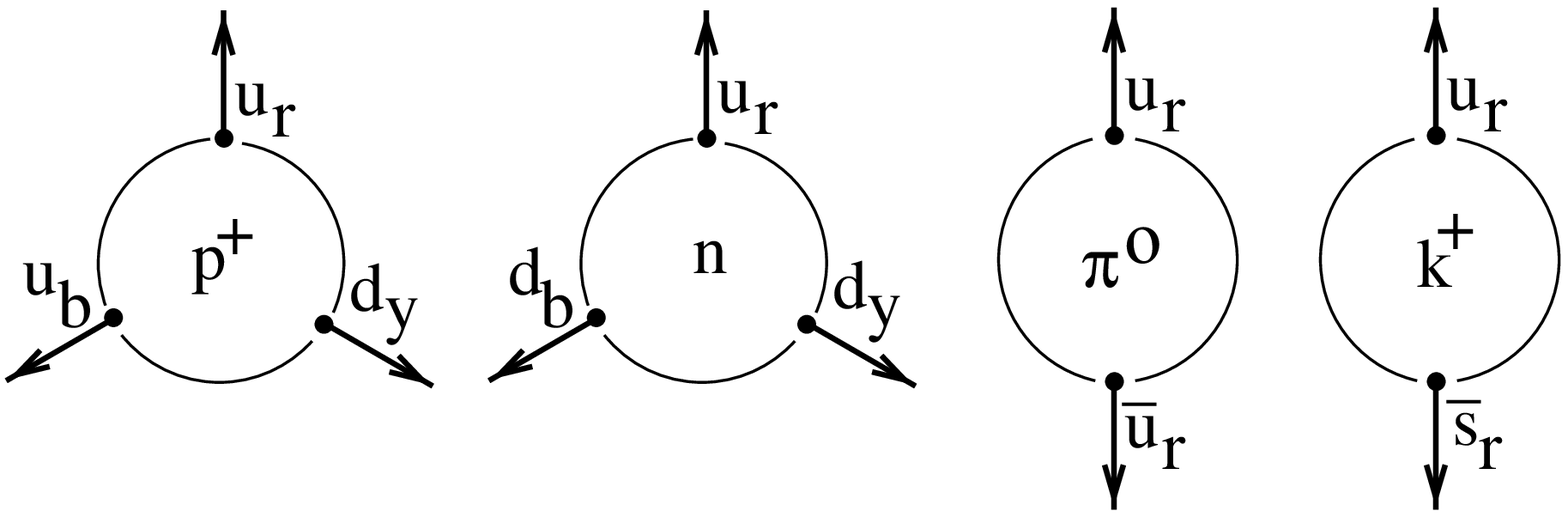} 
\label{fig:9959fig3}
\begin{center}
Fig. 3. The quark structure of proton, nutron, $\pi$ and $K$ mesons
along with their colour charges (the bar denotes antiparticles).
\end{center}

A dramatic consequence of the nonabelian nature of the QCD is that the
gluons themselves carry colour charge and hence have self-interaction
unlike the photons, which have no electric charge and hence no
self-interaction.  Because of the gluon self-interaction the colour
lines of forces between the quarks are squeezed into a tube as
illustrated below.

\vspace{4cm}
\includegraphics{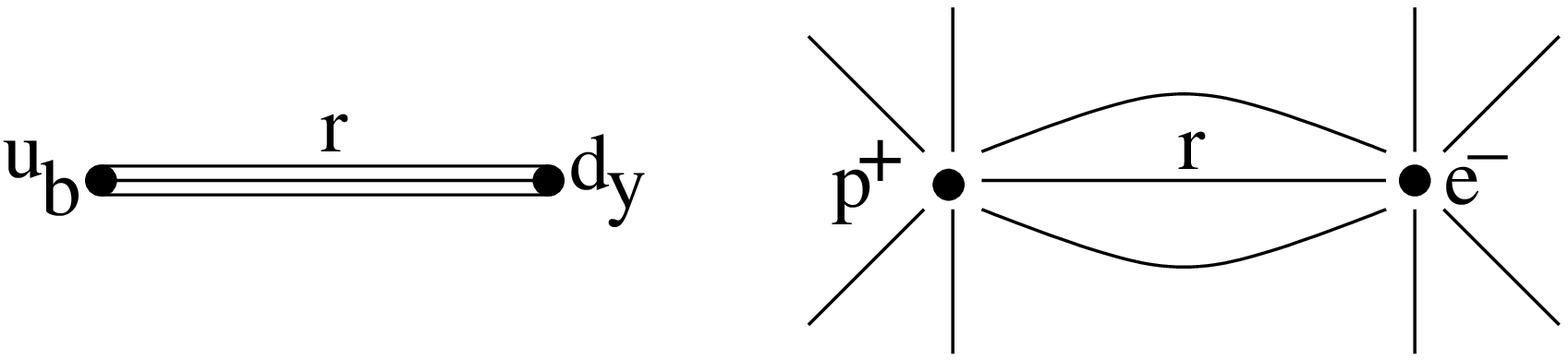} 
\label{fig:9959fig4}
\begin{center}
Fig. 4. The squeezed (1-dim.) lines of force between colour charges
contrasted with the isotropic (3-dim.) lines of force between electric
charges.
\end{center}

\noindent Consequently the number of colour lines of force intercepted
and the resulting force is constant, i.e. the potential increases
linearly with distance
\be
V_s = \alpha_s r.
\label{four}
\ee
Thus the quarks are perpetually confined inside the hadrons as it
would cost an infinite ammount of energy to split them apart.  In
contrast the isotropic distribution of the electric lines of force
implies that the number intercepted and hence the resulting force
decreases like $1/r^2$, i.e.
\be
V_E = {\alpha \over r}.
\label{five}
\ee

Finally the weak interaction is mediated by massive vector particles,
the charged $W^\pm$ and the neutral $Z^0$ bosons, which couple to all
the quarks and leptons.  The former couples to each pair of quarks and
leptons listed above with a universal coupling strength $\alpha_W$,
since they all belong to the doublet representation of $SU(2)$
(i.e. carry the same gauge charge).  This is illustrated in Fig. 2c.
Note the $W$ mass term appearing in the boson propagator, represented
by the denominator of the amplitude.
The Fourrier transform of this quantity gives the weak potential 
\be
V_W = {\alpha_W \over r} e^{-r M_W},
\label{six}
\ee
i.e. the weak interaction is restricted to a short range $\sim
1/M_W$.  One can understand this easily from the uncertainty principle
(\ref{one}), since the exchange of a massive $W$ boson implies a
tansient energy nonconservation $\Delta E = M_W c^2$, corresponding to
a range $\Delta x = \hbar/M_W c$.  Fig. 2c represents the charged
current neutrino scattering processes 
\be
\nu_e \mu^- \rightarrow e^- \nu_\mu, ~~\nu_e d \rightarrow e^- u
\label{seven}
\ee
as well as the corresponding decay processes
\be
\mu^- \rightarrow \nu_\mu e^- \bar\nu_e, ~~ d \rightarrow u e^-
\bar{\nu_e} 
\label{eight}
\ee
since an incoming particle is equivalent to an outgoing antiparticle.
The first process is the familiar muon beta decay while the second is
the basic process underlying neutron beta decay.

The weak and the electromagnetic interactions have been successfully
unified into a $SU(2) \times U(1)$ gauge theory by Glashow, Salam and
Weinberg for which they were awarded the 1979 Nobel Prize.  According
to this theory the weak and the electromagnetic interactions have
similar coupling strengths, i.e.
\be
\alpha_W = \alpha/\sin^2 \theta_W,
\label{nine}
\ee
where $\sin^2 \theta_W$ represents the mixing between the two gauge
groups.  This parameter can be determined from the relative rates of
charged and neutral current ($W^\pm$ and $Z^0$ exchange) neutrino
scattering, giving
\be
\sin^2 \theta_W \simeq 1/4, {\rm i.e.} ~\alpha_W \simeq 1/32.
\label{ten}
\ee
Thus the weak interaction is in fact somewhat stronger than the
electromagnetic.  Its relative weakness in low energy processes like
$\mu$ decay $(Q^2 \ll M^2_W)$ is due to the additional $M^2_W$ term in
the boson propagator, reflecting its short range.  This suppression
is a transient phenomenon, which goes away at a high energy scale
$(Q^2 \gsim M^2_W)$, where the rates of weak and EM interactions
become similar.  A direct prediction of this theory is the mass of the
$W$ boson from the $\mu$ decay amplitude $\alpha_W/M^2_W$ or the
equivalent Fermi coupling $G_F$.  The observed $\mu$ decay rate gives 
\be
G_F \equiv {\pi \over \sqrt{2}} {\alpha_W \over M^2_W} = 1.17 \times
10^{-5} ~{\rm GeV}^{-2}.
\label{eleven}
\ee
From (\ref{ten}) and (\ref{eleven}) we get
\be
M_W \simeq 80 ~{\rm GeV}~{\rm and}~ M_Z = M_W/\cos\theta_W \simeq 91
~{\rm GeV}.
\label{twelve}
\ee
\bigskip

\noindent \underbar{\bf Discovery of the Fundamental Particles}
\medskip

As mentioned earlier, the up and down quarks are the constituents of
proton and neutron.  Together with the electron they constitute all
the visible matter around us.  The heavier quarks and charged leptons
all decay into the lighter ones via interactions analogous to the muon
decay (\ref{eight}).  So they are not freely occurring in nature.  But
they can be produced in laboratory or cosmic ray experiments.  The
muon and the strange quark were discovered in cosmic ray experiments
in the late forties, the latter in the form of $K$ meson.  Next to
come were the neutrinos.  Although practically massless and stable the
neutrinos are hard to detect because they interact only weakly with
matter.  The $\nu_e$ was discovered in atomic reactor experiment in
1956, for which Reines got the Nobel Prize in 1995.  The $\nu_\mu$ was
discovered in the Brookhaven proton synchrotron in 1962, for which
Lederman and Steinberger got the Nobel Prize in 1988.  The first
cosmic ray observation of neutrino came in 1965, when the $\nu_\mu$
was detected in the Kolar Gold Field experiment. 

The rest of the particles have all been discovered during the last 25 years,
thanks to the advent of the electron-positron and the
antiproton-proton colliders.  First came the windfall of the seventies
with a quick succession of discoveries mainly at the $e^+e^-$
colliders: charm quark (1974), Tau lepton (1975), bottom quark (1977)
and the gluon (1979).  This was followed by the discovery of $W$ and
$Z$ bosons (1983) and finally the top quark (1995) at the $\bar pp$
colliders. In view of the crucial role played by the colliders in
these discoveries a brief discussion on them is in order.
\bigskip

\noindent \underbar{\bf The $e^+e^-$ and $\bar pp$ Colliders}
\medskip

These are synchrotron machines, where the particle and antiparticle
beams are simultaneously accelerated in the same vacuum pipe using the
same set of bending magnets and accelerating field.  Thanks to their
equal mass and opposite charge the particle and antiparticle beams go
around in identical orbits on top of one another throughout the course
of acceleration (Fig. 5a).  On completion of the acceleration mode,
the two beams are made to collide almost head on by flipping a
magnetic switch (Fig. 5b).  In this mode the two beams continue to
collide repeatedly at the collision points.  Indeed, the machine
spends a major part of its running time in the collision mode.

\vspace{5cm}
\includegraphics{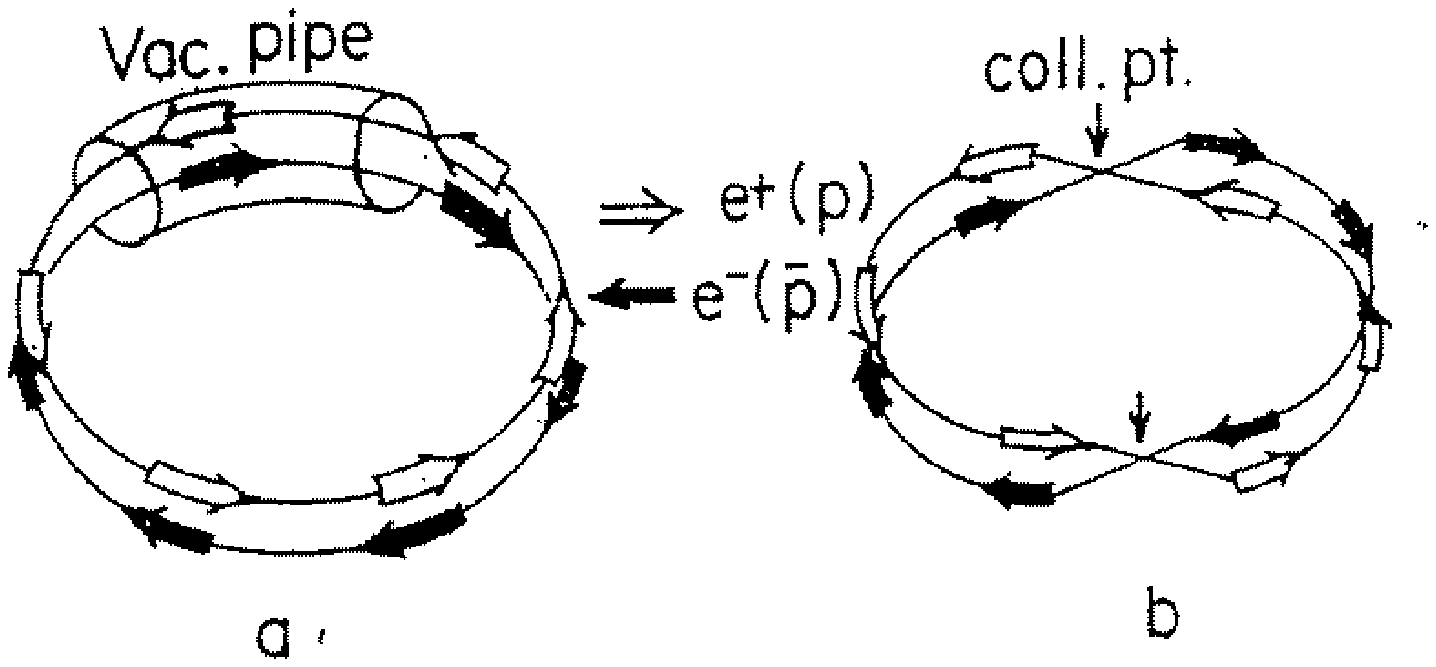} 
\label{fig:9959fig5}
\begin{center}
Fig. 5. The $e^+e^-$ (or $\bar pp$) collider.
\end{center}

The advantage of a collider over a fixed target machine is an enormous
gain of the Lorentz invariant energy $s$ (CM energy squared) at
practically no extra cost.  For, a collider (Fig. 6a) and a fixed
target machine (Fig. 6b) correspond respectively to 
\be
s = (2E)^2 ~{\rm and}~ s = 2mE
\label{thirteen}
\ee
where $E$ denotes the beam energy and $m$ the target particle mass.

Thus a collider of beam energy $E$ is equivalent to a fixed target
machine of beam

\vspace{3cm}
\includegraphics{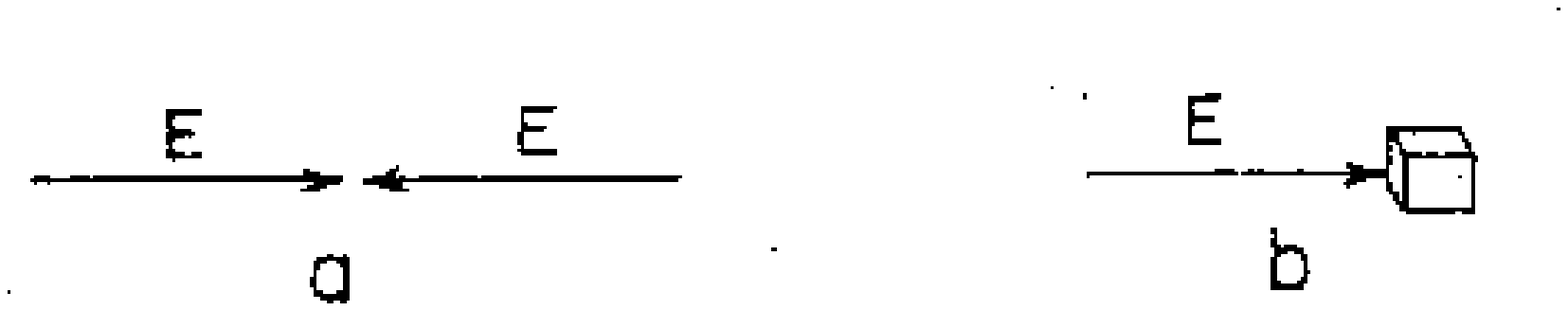} 
\label{fig:9959fig6}
\begin{center}
Fig. 6. Collider vs. Fixed Target Machine
\end{center}

\noindent energy
\be
E' = 2E^2/m.
\label{fourteen}
\ee
Thus the TEVATRON $\bar pp$ collider beam energy of 1 TeV (1000 GeV)
is equivalent to a fixed target machine beam of 2000 TeV, since $m_p
\simeq 1$ GeV.  For a $e^+e^-$ collider, of course, the energy gain is
another factor of a 1000 higher due to the smaller electron mass. 
\medskip

\noindent \underbar{$\bar pp$ vs. $e^+e^-$ Collider} -- A proton
(antiproton) is nothing but a beam of quarks (anti-quarks) and
gluons.  Thus the $\bar pp$ collision can be used to study $\bar qq$
interaction which can probe the same physics as the $e^+e^-$
interaction (Fig. 7).

\vspace{5cm}
\includegraphics{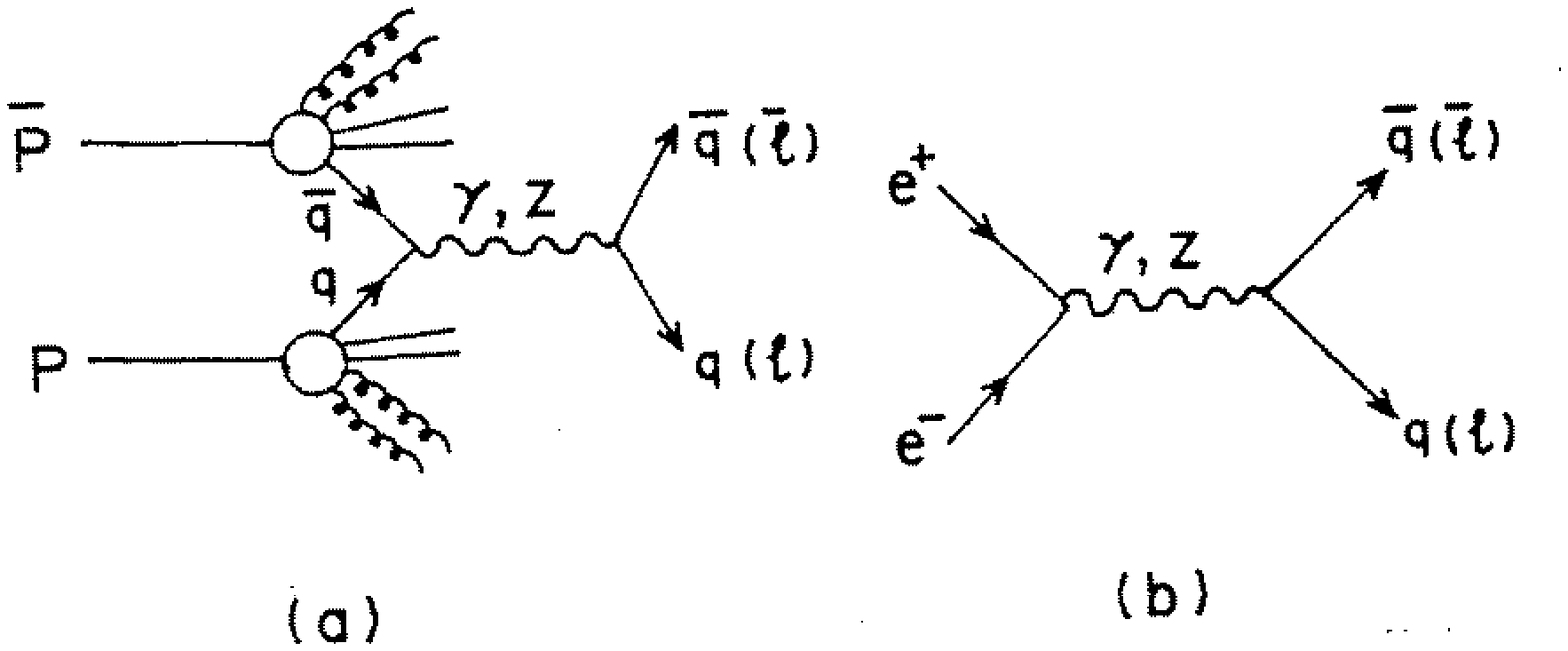} 
\label{fig:9959fig7}
\begin{center}
Fig. 7. $\bar pp$ vs $e^+e^-$ collision
\end{center}

Of course, one has to pay a price in terms of the beam energy, since a
quark carries only about 1/6 of the proton energy\footnote{The proton
energy momentum is shared about equally between the quarks and the
gluons; and since there are 3 quarks ($uud$), each one has a share of
$\sim 1/6$ on the average.}.  Thus
\be
s^{1/2}_{e^+e^-} \sim s^{1/2}_{\bar qq} \sim (1/6) s^{1/2}_{\bar pp}, 
\label{fifteen}
\ee
i.e. the energy of the $\bar pp$ collider must be about 6 times as
large as the $e^+e^-$ collider to give the same basic interaction
energy.  This is a small price to pay, however, considering the
immensely higher synchrotron radiation loss for the $e^+e^-$
collider.  The amount of synchroton radiation loss per turn is
\be
\Delta E \simeq {4\pi \over 3} {e^2 E^4 \over m^4 \rho},
\label{sixteen}
\ee
where $e,m$ and $E$ are the particle charge, mass and energy and
$\rho$ is the radius of the ring.  This quantity is much larger for
the $e^+e^-$ collider due to the small electron mass, which would mean
a collosal energy loss and radiation damage.  These are reduced by
increasing the radius of the ring, resulting in a higher construction
cost.  The point is best illustrated by the following comparison
between the $\bar pp$ collider (S$\bar{\rm P}$PS) and the $e^+e^-$
collider (LEP) at CERN, both designed for producing the $Z$ boson.
\[
\begin{tabular}{llll}
Machine & Energy $(s^{1\over2})$ & Radius ($\rho$) & cost \\
&&& \\
S$\bar{\rm P}$PS ($\bar pp$) & 600 GeV & 1 Km & \$ 300 million \\
&&& \\
LEP $(e^+e^-)$ & 100 GeV & 5 Km & \$ 1000 million \\
\end{tabular}
\]
Of the total construction cost of the $\bar pp$ collider, \$ 200
million went into building the fixed target super proton synchrotron
(SPS) and only about a \$ 100 million in converting it into a $\bar
pp$ collider.

On the other hand, the $e^+e^-$ collider has an enormous advantage
over $\bar pp$ as a tool for precise and detailed investigation.
This is because one can tune the $e^+e^-$ energy to a desired particle
mass (e.g. $M_Z$), which cannot evidently be done with the
quark-antiquark energy.  Thus one could produce about a million of $Z
\rightarrow e^+e^-$ events per year at LEP while the annual yield of
such events at the S$\bar{\rm P}$PS was only about a few dozen.  Moreover
the $e^+e^-$ collider signals are far cleaner than the $\bar pp$
collider since one has to contend with the debris from the remaining
quarks and gluons in the latter case.

In short the $\bar pp$ collider is more suitable for surveying a new
energy domain, being comparatively less expensive, while the $e^+e^-$
is better suited for intensive follow-up investigation.  The following
table gives a list of the important colliders starting from the early
seventies.  

First came the Stanford $e^+e^-$ collider (SPEAR) with a CM energy
similar to that of the fixed-target proton syncrotron at Brookhaven.
In fact the charm quark was simultaneously discovered at both these
machines in 1974, for which Richter and Ting got the 1975 Nobel
Prize.  The tau lepton was also discovered at SPEAR the following
year, for which Pearl got the Nobel Prize in 1995.  The bottom quark
was discovered in the fixed-target proton synchrotron at Fermilab in
1977; and soon followed by the study of its detailed properties at the
$e^+e^-$ colliders at Hamburg (DORIS) and Cornel (CESR).  Then the
construction of a more energetic $e^+e^-$ collider (PETRA) at Hamburg
resulted in the discovery of gluon in 1979.
\newpage
\[
\begin{tabular}{|l|c|c|c|c|c|}
\multicolumn{6}{c}{Table 1. Past, Present and Proposed Colliders} \\
\hline
Machine & Location & Beam & Energy & Radius & Highlight \\
        &          &      & (GeV)  &        &           \\
\hline
SPEAR & Stanford & $e^+e^-$ & $3+3$ & & $c,\tau$ \\
&&&&& \\
DORIS & Hamburg & '' & $5+5$ & & $b$ \\
&&&&& \\
CESR & Cornell & '' & $8+8$ & 125 m & '' \\
&&&&& \\
PEP & Stanford & '' & $18+18$ & & -- \\
&&&&& \\
PETRA & Hamburg & '' & $22+22$ & 300 m & $g$ \\
&&&&& \\
TRISTAN & Japan & '' & $30+30$ & & -- \\
&&&&& \\
S$\bar{\rm P}$PS & CERN & $\bar pp$ & $300+300$ & 1 Km & $W,Z$ \\
&&&&& \\
TEVATRON & Fermilab & '' & $1000+1000$ & & $t$ \\
&&&&& \\
SLC & Stanford & $e^+e^-$ & $50+50$ & -- & $Z$ \\
&&&&& \\
LEP-I & CERN & '' & '' & 5 Km & $Z$ \\
&&&&& \\
(LEP-II) & & '' & $100+100$ & '' & $W$ \\
&&&&& \\
HERA & Hamburg & $ep$ & $30+800$ & 1 Km & -- \\
&&&&& \\
LHC & CERN & $pp$ & $7,000+7,000$ & 5 Km & Higgs? \\
(LEP Tunnel) & & & & & SUSY? \\
&&&&& \\
NLC & ? & $e^+e^-$ & $500+500$? & -- & '' \\
\hline
\end{tabular}
\]

The charm, bottom and tau production via Fig. 7b results in back to
back pairs of particle and antiparticle, as illustrated in Fig. 8.
The typical life time of these particles are $\sim 10^{-12}$ sec,
corresponding to a range of $c\tau \sim 300 \mu m (0.3 mm)$ at
relativistic energies.  This is adequate to identify these particles
before their decay using high resolution silicon vertex detectors.
The production of gluon can be infered from the observation of 3-jet
events as one of the quark-antiquark pair produced via Fig. 7b
radiates an energetic gluon.  How are the coloured quarks and gluons
able to come out of the confinement region?  This is possible because
quantum mechanical vacuum contains many quarks and gluons, with the
uncertainty principle accounting for their mass and kinetic energy.
Each of the produced particles picks up some of these extra quarks and
gluons to come out as a colourless cluster of hadrons, sharing its
original momentum.  There is only a limited momentum spread 
due to the intrinsic momenta of these `vacuum particles' --
i.e. $\Delta p \sim 0.2$ GeV corresponding to a confinement range of
$\sim 1$ fm.  Thus the produced quarks and gluons can each be
recognised as a narrow and energetic jet of hadrons. 

\vspace{8cm}
\includegraphics{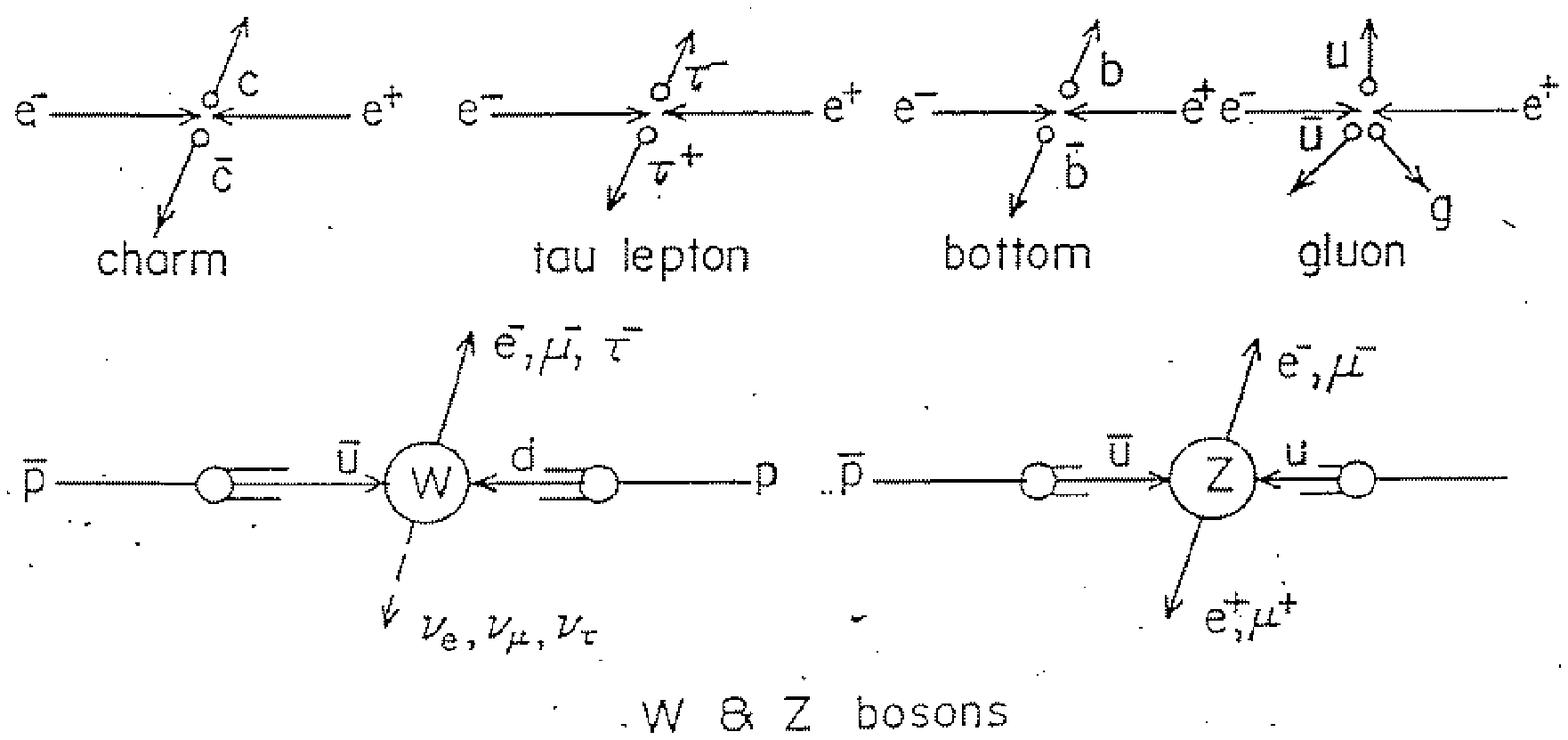} 
\label{fig:9959fig8}
\begin{center}
Fig. 8. 
\end{center}

Next came the CERN $\bar pp$ collider, with a higher CM energy of 100
GeV, even after accounting for the above mentioned factor of 6.  This
resulted in the discovery of $W$ and $Z$ bosons in 1983, for which
Rubbia got the Nobel Prize the following year.  The construction of
the Fermilab $\bar pp$ collider (TEVATRON) increased this effective CM
energy further to $\sim 300$ GeV and resulted in the discovery of the
top quark in 1995.  Being very heavy, these particles decay almost at
the instant of their creation.  Nonetheless they can be recognised by
the unmistakable imprints they leave behind in their decay products,
as illustrated above for the $W$ and $Z$ decays (Fig. 8).  The huge
energy released in the $W \rightarrow \ell \nu$ decay often results in
a hard lepton, carrying a transverse momentum $(p_T) \sim M_W/2 \sim
40$ GeV, with an apparent $p_T$-inbalance (missing-$p_T$) as the
neutrino escapes detection.  Similarly the $Z \rightarrow \ell^+
\ell^-$ decay results in a pair of azimuthally back-to-back hard
leptons, carring a $p_T \sim M_Z/2 \sim 45$ GeV each.  The top quark
event is more complex, since they are produced in pair and each
undergoes a 3-body decay via $W$ as shown in Fig. 9.  This
results in a hard isolated lepton and a number of hard quark jets.  The lepton
isolation and the jet hardness criteria have been successfully
exploited to extract the top quark signal from the background.  

\newpage

\hrule width 0pt
\includegraphics{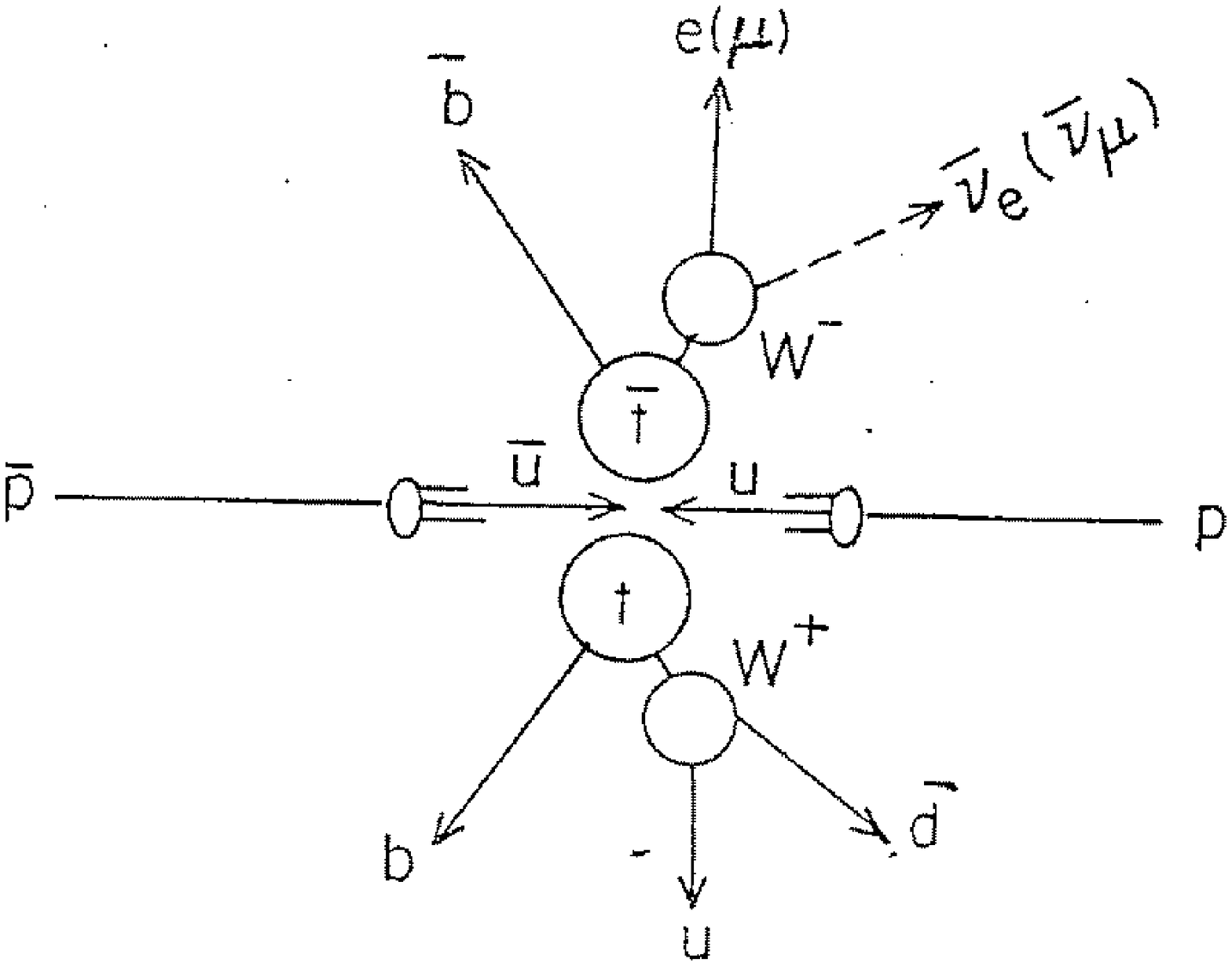} 
\label{fig:9959fig9}
\vspace{9cm}
\begin{center}
Fig. 9.
\end{center}

The last batch of machines are the Stanford linear collider (SLC) and
the large electron-positron collider (LEP).  They have helped to
bridge the energy gap between the $e^+e^-$ and the $\bar pp$ colliders
and made a detailed study of $Z$ and $W$ bosons.  Moreover a $ep$
collider (HERA) at Hamburg is probing deeper into the structure of the
proton by extending the $ep$ scattering experiment of Fig. 1 to higher
energies.  

Finally a large hadron collider (LHC), being built in the LEP tunnel
at a cost of 3-4 billion dollars,
is expected to be complete by 2005.  It will have a total CM energy of
14 TeV, i.e. an effective energy of 2 TeV (2000 GeV).  Thus it will
push up the energy frontier by almost an order of magnitude.  The
reason for going to these higher energies is that, although we have
seen all the basic constituents of matter and the carriers of their
interactions, the picture is not complete yet.  As we shall see below,
a consistent theory of their masses requires the presence of a host of
new particles in the mass range of a few hundred GeV, which await
discovery at the LHC.  Indeed there is already an active proposal for
a detailed follow-up investigation of this energy range with a
$e^+e^-$ machine, generically called the next linear collider (NLC),
although one does not know yet when and where.
\bigskip

\noindent \underbar{\bf Mass Problem (Higgs Mechanism)}
\medskip

By far the most serious problem with the Standard Model is how to give
mass to the weak gauge bosons (as well as the quarks and leptons)
without breaking the gauge symmetry of the Lagrangian, which is
essential for a renormalisable field theory.  Let us illustrate this
with the simpler example of a $U(1)$ gauge theory, describing the EM
interaction of a charged scalar (spin 0) field $\phi$.  The
corresponding Lagrangian is
\beq
{\cal L}_{EM} &=& (\partial_\mu - ieA_\mu) \phi^\star(\partial_\mu +
ieA_\mu) \phi - [\mu^2 \phi^\star \phi - \lambda(\phi^\star \phi)^2] -
{1\over4} F_{\mu\nu} F_{\mu\nu}, \nonumber \\[2mm] 
F_{\mu\nu} &=& \partial_\mu A_\nu - \partial_\nu A_\mu,
\label{seventeen}
\eeq
where $A_\mu$ is the EM field (photon) and $F_{\mu\nu}$ the
corresponding field tensor.  Since $\partial_\mu$ is the momentum
operator the last term of the Lagrangian represents the gauge kinetic
energy.  The middle term represents the scalar mass and
self-interaction, while the first one represents scalar kinetic energy
and gauge interaction.  It is clear that the scalar mass and self
interaction terms are invariant under the phase transfermation 
\be
\phi \rightarrow e^{i\alpha(x)} \phi,
\label{eighteen}
\ee
which is called gauge transformation for historical reason.  But
$\partial_\mu \phi$ and the resulting scalar kinetic energy term are
evidently not invariant under this phase (gauge) transformation.
However the invariance of the Lagrangian under this gauge
transformation on $\phi$ is restored if one makes a simultaneous
transformation on $A_\mu$, i.e.
\be
A_\mu \rightarrow A_\mu + {1 \over e} \partial_\mu \alpha(x).
\label{ninteen}
\ee
Thus it is a remarkable property of gauge theory that the symmetry of
the Lagrangian under the phase transformation of a charged particle
field implies its interaction dynamics.  But the symmetry of the
Lagrangian would be lost once we add a mass term for the gauge field,
$M^2 A_\mu A_\mu$, which is clearly not invariant under
(\ref{ninteen}).  While the photon has no mass the analogous gauge
bosons for weak interaction, $W$ and $Z$, are massive.  So the
question is how to add such mass terms without destroying the gauge
symmetry of the Lagrangian.  

The clue is provided by the observation that the scalar mass term,
$\mu^2 \phi^\star \phi$, is gauge symmetric.  This is exploited to
give mass to $W$ and $Z$ (as well as the quarks and leptons)
through backdoor.  They acquire mass by absorbing scalar particles,
somewhat similar to a snake acquiring mass by swallowing a rabbit.
This is the famous Higgs mechanism.  One starts with a scalar field of
imaginary mass (negative $\mu^2$).  The corresponding scalar potential
\be
V = \mu^2 \phi^\star \phi + \lambda (\phi^\star \phi)^2
\label{twenty}
\ee
is shown in the 3-dimensional plot below (Fig. 10a), as a function of
the complex field.

\vspace{6cm}
\includegraphics{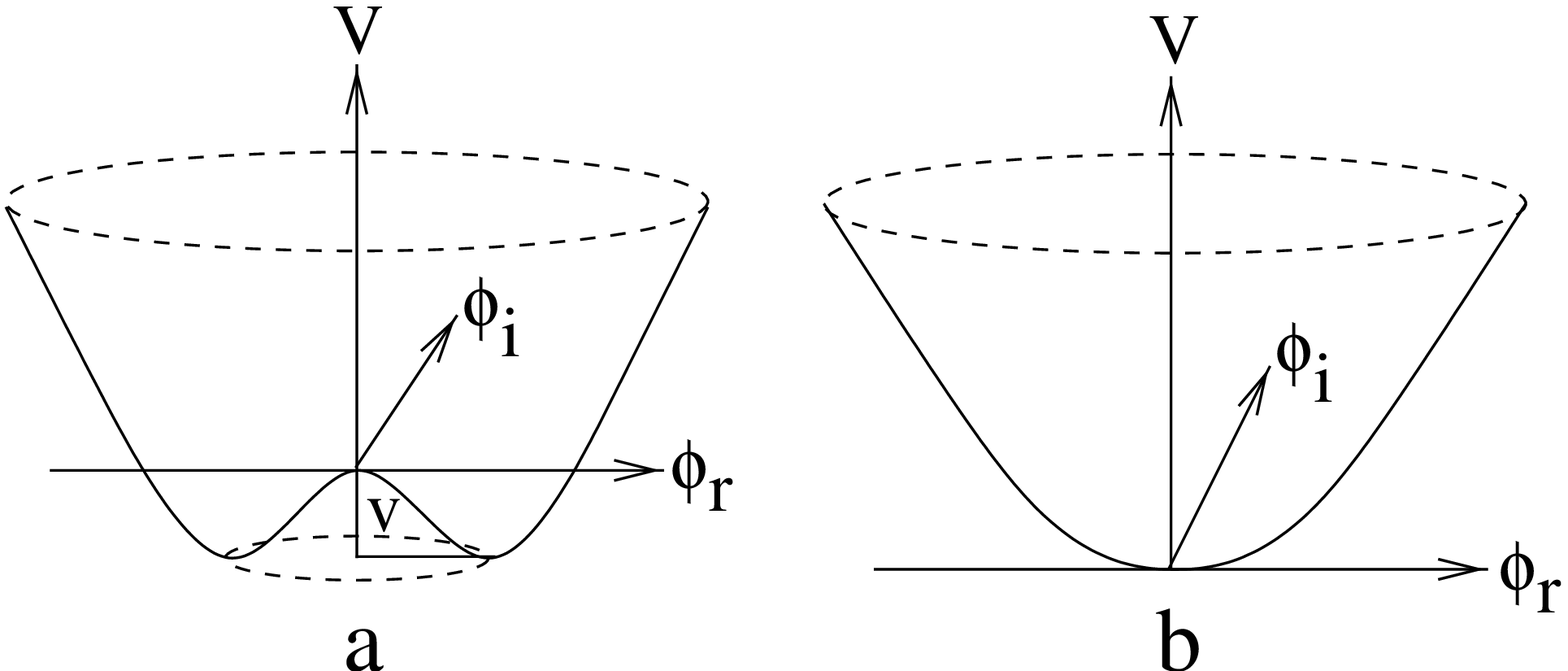} 
\label{fig:9959fig10}
\begin{center}
Fig. 10.  The scalar potential (\ref{twenty}) as a function of the
complex scalar field with (a) negative $\mu^2$ and (b) positive
$\mu^2$, resembling the bottom of a wine bottle and the top of a wine
glass respectively.   
\end{center}

\noindent Starting from zero it becomes negative at small
$|\phi|$ where the quadratic term wins and positive at large $|\phi|$
where the quartic term takes over.  The minimum occurs at a finite
value of $|\phi|$, i.e. at
\be
v = \sqrt{-\mu^2/2\lambda}.
\label{twentyone}
\ee
Thus the ground state (vacuum state) corresponds to this finite value
of $|\phi|$. 

Note that the potential of Fig. 10a looks like the bottom of a wine
bottle.  It is a useful analogy since the shape of the bottle maps the
potential energy.  A drop of wine left in the bottle will not sit at
the origin but drop off to the rim because it corresponds to minimum
potential energy (ground state).  This may be contrasted with the
potential with positive $\mu^2$, shown in Fig. 10b, which looks like
the top of a wine glass.  In this case the drop of wine will rest at
the origin because it is the state of minimum potential energy.  One
can shake the drop a little and study its motion perturbatively as a
small oscillation problem around the origin, which is the position of
stable equilibrium.  One can do the same thing in the previous case
(Fig. 10a) except that the perturbative expansion has to be done not
around the origin but a point on the rim, because the latter
corresponds to minimum potential energy (stable equilibrium).

It is the same thing in perturbative field theory.  Perturbative
expansions must be done around a point of minimum energy (stable
equillibrium).  So one must translate the origin to this point and do
perturbation theory with the redefined field
\be
h(x) = \phi (x) - v,
\label{twentytwo}
\ee
which represents the physical scalar (called the Higgs boson).
Substituting (\ref{twentytwo}) in (\ref{seventeen}) we find several
pleasant surprises.  The first term of the Lagrangian gives a
quadratic term in $A^\mu$, $e^2 v^2 A_\mu A_\mu$, corresponding to a
mass of the gauge boson 
\be
M = ev.
\label{twentythree}
\ee
Moreover the middle term leads to a real mass for the physical scalar,
\be
M_h = \sqrt{-\mu^2} = (\sqrt{2\lambda})v = {\sqrt{2\lambda} \over e}
M.
\label{twentyfour}
\ee
Thus the Higgs boson mass is equal to the mass of the gauge boson times the
ratio of the self-coupling to the gauge coupling.  Although we do not
know the value of the self-coupling $\lambda$, the validity of
perturbation theory implies $\lambda \lsim 1$.  This means that the
Higgs boson mass is expected to be in the same ball park as $M_W$,
i.e. within a few hundred GeV.  Hence it is a prime candidate for
discovery at the LHC.

In order to appreciate the Higgs mechanism let us look back at
Fig. 10.  The choice of negative $\mu^2$ meant that the ground state
has moved from the origin to a finite value of $|\phi|$ (i.e. a point
on the rim).  While the former was invariant under phase
transformation, the latter point is not.  Thus the phase (gauge)
symmetry of the Lagrangian is not shared by the ground state
(vacuum).  This phenomenon is known as spontaneous symmetry breaking.
There are many examples of this in physics, the most familiar one
being that of magnetism.  As we cool a Ferromagnet below a critical
temperature the electron spins get alligned with one another because
that corresponds to ground state of energy.  Thus the Lagrangian
possesses a rotational symmetry, but not the ground state.  The same
thing happens in the Higgs mechanism, except that the rotation is in
the phase space of the complex field $\phi$ instead of the ordinary
space.  The breaking of the phase (gauge) symmetry by the ground state
enables the gauge boson to acquire mass.  At the same time the fact
that this symmetry is retained by the Lagrangian enables the
renormalisation theory to go through.  Needless to say that this last
feature is very important because it ensures a systematic cancellation
of infinites, without which there would be no predictive field
theory.  The renormalisability of the Electroweak gauge theory in the
presence of spontaneous symmetry breaking was shown to be valid by t'
Hooft in the early seventies, for which he got the Nobel Prize this
year along with his teacher, Veltman.
\bigskip

\noindent \underbar{\bf Hierarchy Problem (Supersymmetry)}
\medskip

The Higgs solution to the mass problem is not the full story because
it leads to the so called hierarchy problem -- i.e. how to control the
Higgs boson mass in the desired range of $\sim 10^2$ GeV.  This is
because the scalar particle mass has quadratically divergent quantum
corrections unlike the fermion or gauge boson masses.  The quantum
corrections arise from the interaction of the particle with those
present in the quantum mechanical vacuum -- i.e. it is analogous to
the effect of medium on a particle mass in classical mechanics.  This
is illustrated by the Feynman diagrams of Fig. 11 below, showing the
radiative loop contributions to the Higgs mass coming from its quartic
self-interaction of (\ref{seventeen}) and its interaction with a
fermion pair.  Thanks to the uncertainty principle, the momentum $k$
of the vacuum particles can be arbitrarily large.  

\vspace{4cm}
\includegraphics{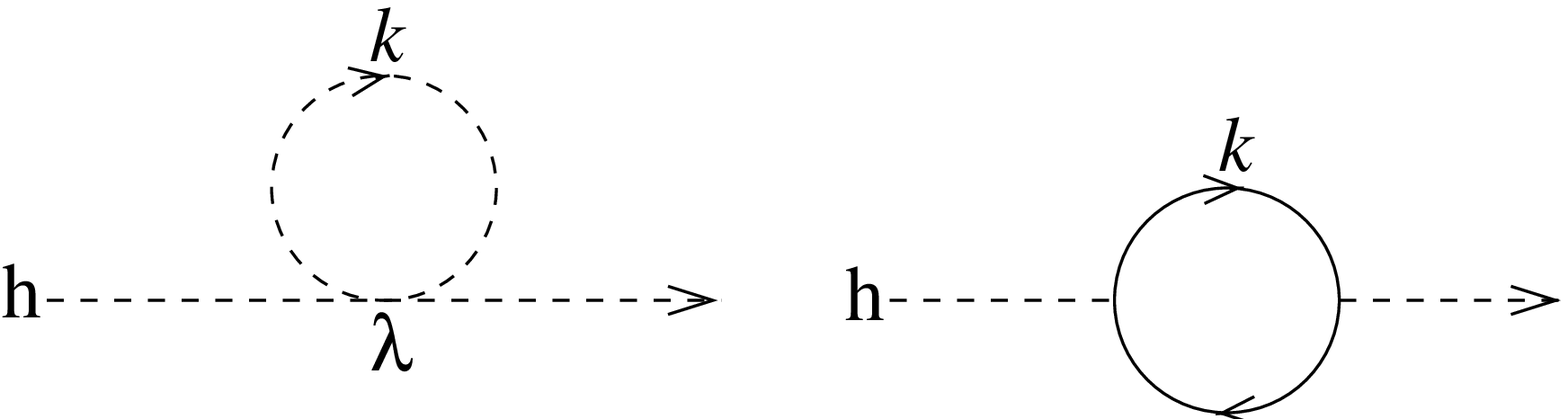} 
\label{fig:9959fig11}
\begin{center}
Fig. 11.  Radiative loop contributions to Higgs boson mass coming from
its (a) quartic self-coupling and (b) coupling to fermion-antifermion pair.
\end{center}

\noindent Integrating the boson propagator factor of Fig. 11a, $1/(k^2
- M^2)$, over this 4-momentum gives 
\be
\int {d^4k \over k^2 - M^2} \sim \int {k^3 dk \over k^2 - M^2} \sim
\int kdk \sim k^2 \Bigg|^\wedge_0,
\label{twentyfive}
\ee
where $\wedge$ is the cut-off scale of the theory.  Similarly
intergrating the product of the fermion and anti-fermion propagators
in Fig. 11b gives
\be
\int {d^4k \over (/\!\!\! k - m)^2} \sim k^2\Bigg|^\wedge_0,
\label{twentysix}
\ee
where $/\!\!\! k = k_\mu \gamma_\mu$, i.e. the invariant product of the
4-momentum with the Dirac matrices.  It may be noted here that an
analogous radiative contribution from fermion loop is also present for
the photon field $A_\mu$.  However there is mutual cancellation
between the divergent contributions to different spin states of the
photon via the so called gauge condition.  In other words it is the
gauge symmetry that protects the gauge boson masses from divergent
quantum corrections.  Similarly the fermion masses are protected by
chiral symmetry.  In the absence of any protecting symmetry the
quadratically divergent quantum corrections from (\ref{twentyfive})
and (\ref{twentysix}) would push up the output Higgs boson mass to the
cutoff scale $\wedge$.  The cutoff scale of the Electroweak theory is
where it encounters new interactions -- typically the GUT scale of
$10^{16}$ GeV or Plank scale of $10^{19}$ GeV.  The former represents
the energy scale where the strong and Electroweak interactions are
presumably unified as per Grand Unified Theory, while the latter
represents the energy where gravitational interaction becomes strong
and can no longer be neglected.  So the question is how to control the
Higgs boson mass in the desired range of $\sim 10^2$ GeV, which is
tiny compared to these cutoff scales.

That the scalar mass is not protected by any symmetry was of course
used in the last section to give mass to gauge bosons and fermions via
Higgs mechanism.  The hierarchy problem encountered now is the flip
side of the same coin.  The most attractive solution is to invoke a
protecting symmetry -- i.e. the supersymmetry (SUSY), which is a
symmetry between fermions and bosons.  As per SUSY all the fermions of
the Standard Models have bosonic Superpartners and vice versa.  They
are listed below along with their spin, where the Superpartners are
indicated by tilde.  
\[
\begin{tabular}{|cc| cc|cc|}
\hline
quarks \& leptons & $S$ & Gauge bosons & $S$ & Higgs & $S$ \\
\hline
&&&&& \\
$q,\ell$ & 1/2 & $\gamma,g,W,Z$ & 1 & $h$ & 0 \\
&&&&& \\
$\tilde q,\tilde \ell$ & 0 & $\tilde\gamma,\tilde g,\tilde W,\tilde Z$
& 1/2 & $\tilde h$ & 1/2 \\ 
&&&&& \\
\hline
\end{tabular}
\]
SUSY ensures cancellation of quadratically divergent contributions
between superpartners, e.g. between the higgs loop of Fig. 11a and the
corresponding fermionic $(\tilde h)$ loop of Fig. 11b.  For the
cancellation to occur to the desired accuracy of $\sim 10^2$ GeV the
mass difference between the superpartners must be restricted to this
scale.  Thus one expects a host of new particles in the mass range of
a few hundred GeV, which can be discovered at the LHC.
\bigskip

\noindent \underbar{\bf Conclusion}
\medskip

The Higgs and the SUSY particles are the minimal set of missing
pieces, required to complete the picture of particle physics.
Therefore the search for these particles is at the forefront of
present and proposed research programmes in this field.  A
comprehensive search for these particles upto the predicted mass limit
of $\sim 1000$ GeV will be possible at the LHC, which should go on
stream in 2005.  Hopefully the LHC will discover these particles and
complete the picture a la the Minimal Supersymmetric Standard Model or
else provide valuable clues to an alternate picture. 

It should be emphasised of course that, while the Supersymmetric
Standard Model represents a complete and self-consistent theory, it is
far from being the ultimate theory.  The ultimate goal of particle
physics is the unification of all interactions.  This has inspired
theorists to propose more ambitious theories unifying all the three
gauge interactions (Grand Unified Theory) and even roping in gravity
(Superstring Theory).  As indicated above, however, the energy scales
of these unifications are many orders of magnitude higher than what
can be realised at present or any foreseeable future experiment.  Thus it
is possible and in the light of our past experience very likely that
nature has many superprises in store for us in the intervening energy
range; and the shape of the ultimate theory (if any) would have little
resemblance to those being postulated today.

\end{document}